\documentclass[a4paper]{jpconf}
\usepackage{graphicx}
\begin{document}
\title{Effects of Radiation on Primordial Non-Gaussianity}

\author{Suratna Das}
\address{Tata Institute of Fundamental Research, Mumbai 400005, India}
\ead{suratna@tifr.res.in}

\begin{abstract}
We study the non-Gaussian features in single-field slow-roll
inflationary scenario where inflation is preceded by a radiation
era. In such a scenario both bispectrum and trispectrum
non-Gaussianities are enhanced. Interestingly, the trispectrum in this
scenario does not depend up on the slow-roll parameters and thus
$\tau_{NL}$ is larger than $f_{NL}$ which can be a signature of such a
pre-inflationary radiation era.
\end{abstract}
\section{Introduction}

Study of non-Gaussian features in primordial perturbations generated
during inflation has become a subject of great importance as the
precise determination of these primordial non-Gaussianities can
quantify the dynamics of the early universe \cite{matarrese}. In
generic single-field slow-roll inflationary scenario the preferred
initial vacuum chosen for the inflaton perturbations is the
Bunch-Davies vacuum. It is shown in \cite{prl1}, that if inflation is
preceded by a radiation era then the inflaton fluctuations will have
an initial thermal distribution where the initial vacuum will depart
from the standard Bunch-Davies one. The presence of pre-inflationary
radiation era enhances the power spectrum of scalar modes by an extra
temperature depended factor $\coth(k/2T)$. The enhanced power spectrum
is in accordance with the observations if the comoving temperature $T$
of the primordial perturbations is less than $10^{-3}$ Mpc$^{-1}$
\cite{prl1}. In this talk we will show that presence of
pre-inflationary radiation era not only enhances the power spectrum
but also generates large bispectrum and trispectrum and these
non-Gaussianities will carry signatures of such pre-inflationary
radiation era \cite{suratna} which will be discussed in detail.
\section{Bispectrum and Trispectrum in single field slow-roll inflation}

The derivations shown in this talk are done in {\it spatially flat
  gauge}. This gauge is preferred in the derivations as in this gauge
the comoving curvature perturbation ${\mathcal R}(t,{\mathbf x})$ is
proportional to the inflaton fluctuations $\delta\phi(t,{\mathbf x})$
as
\begin{eqnarray}
{\mathcal R}(t,{\mathbf x})=\frac{H}{\dot{\phi}}\delta\phi(t,{\mathbf x}),
\end{eqnarray}
where $H$ is the Hubble parameter and the overdot represents
derivative w.r.t. cosmic time $t$. Thus, in this gauge, the comoving
curvature power spectrum, i.e. the two-point correlation function of
${\mathcal R}(t,{\mathbf k})$ in Fourier space, is directly related to
inflaton's power spectrum as
\begin{eqnarray}
{\cal P}_{\cal R}(k)=\frac{k^3}{2\pi^2}\langle{\mathcal R}(k){\mathcal R}(k)\rangle\longleftrightarrow\left(\frac{H}{\dot{\phi}}\right)^2\langle\delta\phi(k)\delta\phi(k)\rangle.
\end{eqnarray}
The comoving curvature power spectrum is measured through observations
of the $TT$ anisotropy spectrum of CMBR which is nearly
scale-invariant.
 
Bispectrum, the non-vanishing three-point correlation function of
primordial fluctuations, is the lowest order departure from
Gaussianity of those primordial perturbations. The non-Gaussianity
arising from bispectrum is quantified by a non-linear parameter,
$f_{NL}$, which is constrained by several experiments as : (i) the
WMAP 5yr data yields $-151 < f_{NL}^{eq}<253$ ($95 \%$ CL)
\cite{Komatsu}, (ii) the PLANCK mission is sensitive to probe
bispectrum upto $f_{NL} \sim 5$ \cite{spergel} and (iii) in future
experiments if the primordial non-Gaussianities imprinted in 21 cm
background is measured then $f_{NL}< 0.1$ can be probed
\cite{cooray,cooray-prd}. In a free theory, as the primordial
perturbations are Gaussian in nature, the three-point correlation
function vanishes yielding no non-Gaussianity. It is shown in
\cite{raghu} that self-interactions of inflaton field of the kind
$V(\phi)=\lambda\phi^3$ generates non-vanishing bispectrum
proportional to $\lambda/H$ but the non-Gaussianity is too small
$(\sim\mathcal{O}(10^{-7}))$ to be probed by any existing or future
experiments. On the other hand, in a generic single-field slow-roll
inflationary model non-linearities in the evolution of primordial
perturbations ${\mathcal R}(t,{\mathbf k})$ can also generate
primordial non-Gaussianities in CMBR. In the non-linear limit one can
write
\begin{eqnarray}
{\mathcal R}_{NL}(t,{\mathbf x})=\frac{H}{\dot{\phi}}\delta\phi_L(t,{\mathbf x})+\frac12\frac{\partial}{\partial\phi}\left(\frac{H}{\dot{\phi}}\right)\delta\phi_L^2(t,{\mathbf x})+{\mathcal O}(\delta\phi_L^3),
\label{non-linear}
\end{eqnarray}
which yields a non-vanishing three-point correlation function of
${\mathcal R}_{NL}$ in terms of four-point correlation function of
inflaton perturbations $\delta\phi_L$ as
\begin{eqnarray}
\langle{\mathcal R}_{NL}{\mathcal R}_{NL}{\mathcal R}_{NL}\rangle\simeq\left(\frac{H}{\dot{\phi}}\right)^2\frac12\frac{\partial}{\partial\phi}\left(\frac{H}{\dot{\phi}}\right)\langle\delta\phi_L\delta\phi_L\delta\phi_L^2\rangle,
\label{three-pt}
\end{eqnarray}
even when the initial perturbations $\delta\phi_L$ are Gaussian in
nature. The four-point correlation function on the R.H.S. can be
written in terms of product of two two-point correlation functions and
defining $f_{NL}$ as
\begin{eqnarray}
\left\langle{\mathcal R}({\mathbf k}_1){\mathcal R}({\mathbf k}_2){\mathcal R}({\mathbf k}_3)\right\rangle=(2\pi)^{-\frac32}\delta^3({\mathbf k_1}+{\mathbf k_2}+{\mathbf k_3})\frac65 f_{NL}\left(\frac{P_{\cal R}(k_1)}{k_1^3}\frac{P_{\cal R}(k_2)}{k_2^3}+2\,\,{\rm perms.}\right),
\end{eqnarray}
one can show that the non-linear parameter is of the order of
slow-roll parameters $f_{NL}=\frac56(\delta-\epsilon)$ \cite{wands},
which is also too small $\left(\mathcal{O}(10^{-2})\right)$ to be
detected by any present or forthcoming experiments. The delta-function
in the above equation ensures that the three momenta form a triangle
and $f_{NL}$ is determined in several such triangle configurations,
some of them which we will consider in this talk are : (i) {\it
  Squeezed configuration} $(|{\mathbf k}_1|\approx|{\mathbf
  k}_2|\approx k\gg|{\mathbf k}_3|)$, (ii) {\it Equilateral
  configuration} $(|{\mathbf k}_1|=|{\mathbf k}_2|=|{\mathbf k}_3|=k)$
and (iii) {\it Folded configuration} $(|{\mathbf k}_1|=|{\mathbf
  k}_3|=\frac12|{\mathbf k}_2|=k)$.

The connected part of four-point correlation function of primordial
fluctuations is called the trispectrum
\begin{eqnarray}
\left\langle{\mathcal R}({\mathbf k}_1){\mathcal R}({\mathbf k}_2){\mathcal R}({\mathbf k}_3){\mathcal R}({\mathbf k}_4)\right\rangle_c&\equiv&\left\langle{\mathcal R}({\mathbf k}_1){\mathcal R}({\mathbf k}_2){\mathcal R}({\mathbf k}_3){\mathcal R}({\mathbf k}_4)\right\rangle-\left(\left\langle{\mathcal R}_{L}({\mathbf k}_1){\mathcal R}_{L}({\mathbf k}_2)\right\rangle\left\langle{\mathcal R}_{L}({\mathbf k}_3){\mathcal R}_{L}({\mathbf k}_4)\right\rangle\right.\nonumber\\
&&+2\,\,{\rm perm}\left.\right).
\end{eqnarray}
The non-linear parameter $\tau_{NL}$ quantifies the non-Gaussianity
arising from trispectrum and it is constrained by observations as (i)
WMAP constraints trispectrum as $\left|\tau_{NL}\right|<10^8$
\cite{WMAP-tri}, (ii) PLANCK is expected to reach the sensitivity upto
$\left|\tau_{NL}\right|\sim 560$ \cite{PLANCK-tri} and (iii) future
21cm experiments can probe trispectrum up to the level $\tau_{NL} \sim
10$ \cite{cooray-prd}. A free-scalar theory yields vanishing
trispectrum like vanishing bispectrum. But non-linear evolution of
$\mathcal{R}$ as given in Eq.~(\ref{non-linear}) yields a trispectrum
where $\tau_{NL}=\left(\frac65 f_{NL}\right)^2$ \cite{wands}, which being
proportional to the square of slow-roll parameters is too small to be
detected by any present or future experiments. It is to be noted that
the generic single-field slow-roll inflation predicts a trispectrum
which smaller than the bispectrum by orders of magnitude.

\section{Inflation with prior radiation era and enhanced non-Gaussianity}

If inflation is preceded by a radiation era then the inflaton field
will have an initial thermal distribution where the thermal vacuum
$|\Omega\rangle$ will have finite occupation as
$N_k|\Omega\rangle=n_k|\Omega\rangle$ (the number operator $N_k\equiv
a^\dagger_{\mathbf k}a_{\mathbf k}$). Also there will be a probability
of the system to be in an energy state $\varepsilon_k\equiv n_k k$ as
\begin{eqnarray}
p(\varepsilon_k)\equiv\frac{e^{-\beta n_k k}}{\sum_{n_k}e^{-\beta n_k
    k}}=\frac{e^{-\beta n_k k}}{z}.
\end{eqnarray}
Due to this probability distribution the correlation functions have to
be thermal averaged. Taking into account the initial thermal
distributions of the primordial fluctuations and the probability
distribution due to pre-inflationary radiation era the thermal
averaged inflaton's power spectrum will have an enhancement factor
$1+2f_B(k)$ where $f_B(k)$ is the distribution function of primordial
perturbations. For inflaton (scalar) perturbations the distribution
function will be Bose-Einstein distribution function
$\left(f_B(k)\equiv\frac{1}{e^{\beta k}-1}\right)$ and thus the
enhancement factor will be $1+2f_B(k)=\coth(\beta k/2)$ where
$\beta\equiv\frac1T$ \cite{prl1,suratna}. This enhanced power spectrum
is in accordance with the observations when $T<10^{-3}$ Mpc$^{-1}$
\cite{prl1}.

As the two-point correlation function is thermal averaged due to the
effects of pre-inflationary radiation era, the other higher-point
correlation functions have also to be thermal averaged in a similar
way. For three-point correlation function this will be 
\begin{eqnarray}
\langle{\mathcal R}_{NL}{\mathcal R}_{NL}{\mathcal R}_{NL}\rangle_{\beta}\simeq\left(\frac{H}{\dot{\phi}}\right)^2\frac12\frac{\partial}{\partial\phi}\left(\frac{H}{\dot{\phi}}\right)\langle\delta\phi\delta\phi\delta\phi^2\rangle_{\beta},
\end{eqnarray}
but now the probability of occupancy of a state with four energies
$\epsilon_r$  will be 
\begin{eqnarray}
p(k_1,k_2, k_3,k_4)\equiv\frac{\prod_re^{-\beta n_{k_r}k_r}}{\prod_r\sum_{n_k}e^{-\beta n_{k_r}k_r}}.
\end{eqnarray}
Due to thermal averaging $f_{NL}$ is enhanced. We compute the
non-Gaussianity in this scenario arising from bispectrum in different
triangle configurations : (i) {\it Squeezed configuration} : in this
configuration the enhanced non-Gaussianity is $f_{NL}^{\rm
  th}=f_{NL}\times2\left(1+3.72\coth\left(\frac{\beta
  k}{2}\right)\right)$ where $f_{NL}$ is enhanced by a factor of
64.82, (ii) {\it Equilateral configuration} : in this configuration
the enhanced non-Gaussianity is $f_{NL}^{\rm
  th}=f_{NL}\times\left(3+\frac{5}{4\sinh^2\left(\frac{\beta
    k}{2}\right)}\right)$ where $f_{NL}$ is enhanced by a factor of
90.85, (iii) {\it Folded configuration} : in this configuration the
enhanced non-Gaussianity is $f_{NL}^{\rm
  th}=f_{NL}\times\left(3+\frac{1}{\sinh^2\left(\frac{\beta
    k}{2}\right)}\right)$ where $f_{NL}$ is enhanced by a factor of
73.28.  It can be seen that $f_{NL}$, which is enhanced due to effects
of pre-inflationary radiation era, is within the sensitivity of future
21-cm experiments where primordial non-Gaussianities can be detected
\cite{cooray, cooray-prd}. It is also to be noted that the maximum
non-Gaussianity can arise in the Equilateral configuration when
inflation is preceded by a radiation era. In a later work
\cite{parker} similar analysis is done when perturbations are already
present in the initial vacuum. It is found in \cite{parker} too that
initial presence of quanta in the vacuum can significantly enhance
non-Gaussianity arising from bispectrum which is in agreement with the
results presented here and in \cite{suratna}.

It is very interesting to note at this point that due to thermal
averaging the four-point correlation function is not equal to the
product of two two-point correlation functions which can yield a
non-vanishing connected part as
\begin{eqnarray}
\left\langle{\mathcal R}({\mathbf k}_1){\mathcal R}({\mathbf k}_2){\mathcal R}({\mathbf k}_3){\mathcal R}({\mathbf k}_4)\right\rangle_c&\neq&\left\langle{\mathcal R}({\mathbf k}_1){\mathcal R}({\mathbf k}_2){\mathcal R}({\mathbf k}_3){\mathcal R}({\mathbf k}_4)\right\rangle_\beta\nonumber \\
&-&\left(\left\langle{\mathcal R}_{L}({\mathbf k}_1){\mathcal R}_{L}({\mathbf k}_2)\right\rangle_\beta\left\langle{\mathcal R}_{L}({\mathbf k}_3){\mathcal R}_{L}({\mathbf k}_4)\right\rangle_\beta+2{\rm perm}\right).
\end{eqnarray}
Thus defining the trispectrum in such a situation as
\begin{eqnarray}
\left\langle{\mathcal R}({\mathbf k}_1){\mathcal R}({\mathbf k}_2){\mathcal R}({\mathbf k}_3){\mathcal R}({\mathbf k}_4)\right\rangle_c=\tau_{NL}\left[\frac{P_{\mathcal R}(k_1)}{k_1^3}\frac{P_{\mathcal R}(k_2)}{k_2^3}\delta^3({\mathbf k_1}+{\mathbf k_3})\delta^3({\mathbf k_2}+{\mathbf k_4})+2\,\,{\rm perm.}\right],
\end{eqnarray}
we see that, as the linear perturbations can generate non-vanishing
connected part due to thermal averaging, the non-linear parameter
$|\tau_{NL}|$ will not depend up on slow-roll parameters and can be as
large as 42.58 \cite{suratna} which is within the detection range of
future 21-cm background anisotropy experiments
\cite{cooray-prd}. Hence, we see that the presence of pre-inflationary
radiation era yields larger trispectrum non-Gaussianity than
bispectrum.
 
\section{Conclusion}

The talk was focused on non-Gaussian features in a single-field
slow-roll inflationary model where inflation is preceded by a
radiation era. In a generic single-field slow-roll model of super-cool
inflation non-linear evolution of primordial fluctuations generate
bispectrum non-Gaussianity which is proportional to the slow-roll
parameters \cite{wands} and thus too small to be detected by any
present or future experiments. Non-linear evolution of primordial
fluctuations also generates trispectrum non-Gaussianity where
$\tau_{NL}$ is proportional to the square of slow-roll parameters
\cite{wands}. Thus, this generic inflationary scenario predicts
trispectrum non-Gaussianity which is much smaller than the bispectrum
non-Gaussianity.

We showed that if such a generic inflationary scenario is preceded by
a radiation era it can yield large bispectrum and trispectrum
non-Gaussianities \cite{suratna} which are within the range of
detection of future 21-cm background anisotropy experiments
\cite{cooray, cooray-prd}. Due to presence of pre-inflationary
radiation era the initial vacuum will contain thermal fluctuations and
also the energy states will have a probability
distribution. Accordingly, the thermal averaged three-point
correlation function generates large non-Gaussianity where the
enhancement of $f_{NL}$ over the generic scenario is largest in the
equilateral configuration. An interesting situation arises in the case
of trispectrum as the thermal averaged four-point correlation function
is not equal to the product of two thermal-averaged two-point
correlation functions. Thus the linear primordial perturbations can
generate a non-vanishing connected part of four-point correlation
function due to thermal averaging and $\tau_{NL}$ in such a case will
not depend up on the slow-roll parameters. We compute that in such a
scenario $|\tau_{NL}|$ can be as large as 43 \cite{suratna}. Thus a
significant signature of such pre-inflationary radiation era is that
it yields larger trispectrum than bispectrum. This signature can
distinguish between an inflationary scenario preceded by a radiation
era and the generic scenario of single-field slow-roll super-cool
inflation.

\ack{I thank my collaborator Subhendra Mohanty.}


\end{document}